\documentclass[aps,prm,twocolumn,superscriptaddress,floatfix]{revtex4-1}

\usepackage[latin1]{inputenc}
\usepackage{tabularx}
\usepackage{graphicx}
\usepackage{dcolumn}
\usepackage{bm}
\usepackage{amssymb}
\usepackage{microtype}
\usepackage{xfrac}
\usepackage{makecell}
\usepackage{array}
\usepackage[version=4]{mhchem}
\usepackage{gensymb}
\usepackage{multirow}
\usepackage[colorlinks=true]{hyperref}

\newcommand{\angstrom}{\text{\normalfont\AA}}

\makeatletter
\newcommand\footnoteref[1]{\protected@xdef\@thefnmark{\ref{#1}}\@footnotemark}
\makeatother

\begin{document}
\title{Incommensurate magnetic order drives singular \\ angular magnetoresistance in a Weyl semimetal}

\author{X.~Yao}
\affiliation{Department of Physics, Boston College, Chestnut Hill, MA 02467, USA}

\author{P.~Chen}
\affiliation{NIST Center for Neutron Research, National Institute of Standards and Technology, Gaithersburg, Maryland 20899, USA}

\author{R.~Verma}
\affiliation{Department of Condensed Matter Physics and Materials Science, Tata Institute of Fundamental Research, Colaba, Mumbai 400005, India}

\author{X.~Zhao}
\affiliation{Department of Physics, Boston College, Chestnut Hill, MA 02467, USA}

\author{H.-Y.~Yang}
\affiliation{Department of Physics, Boston College, Chestnut Hill, MA 02467, USA}
\affiliation{Department of Electrical and Computer Engineering, University of California, Los Angeles, CA, USA.}

\author{L.~DeBeer.Schmitt}
\affiliation{Neutron Scattering Division, Oak Ridge National Laboratory, Oak Ridge, Tennessee 37831, USA}

\author{A.~A.~Aczel}
\affiliation{Neutron Scattering Division, Oak Ridge National Laboratory, Oak Ridge, Tennessee 37831, USA}

\author{C.-M.~Wu}
\affiliation{National Synchrotron Radiation Research Center, Hsinchu 300092, Taiwan}

\author{D.~Alba~Venero}
\affiliation{ISIS neutron and muon source, STFC, Rutherford Appleton Laboratory, Didcot, OX11 0QX, UK}

\author{T.~Ohhara}
\affiliation{J-PARC Center, Japan Atomic Energy Agency, 2-4 Shirakata, Tokai, Ibaraki 319-1195, Japan}

\author{K.~Munakata}
\affiliation{Department of Materials Science, Institute of Pure and Applied Sciences, University of Tsukuba, Tsukuba, Ibaraki 305-8573, Japan}

\author{M.~Takahashi}
\affiliation{Department of Materials Science, Institute of Pure and Applied Sciences, University of Tsukuba, Tsukuba, Ibaraki 305-8573, Japan}

\author{Y.~Noda}
\affiliation{Institute for Solid State Physics, The University of Tokyo, Kashiwa, Chiba 277-0882, Japan}
\affiliation{Institute of Multidisciplinary Research fopr Advanced Materials, Tohoku University, Sendai 980-8577, Japan}

\author{A.~Bansil}
\affiliation{Department of Physics, Northeastern University, Boston, Massachusetts 02115, USA}
\affiliation{Quantum Materials and Sensing Institute, Northeastern University, Burlington, Massachusetts 01803, USA}

\author{B.~Singh}
\affiliation{Department of Condensed Matter Physics and Materials Science, Tata Institute of Fundamental Research, Colaba, Mumbai 400005, India}

\author{P.~ Nikoli{\'{c}}}
\affiliation{Institute for Quantum Matter and William H. Miller III Department of Physics and Astronomy, Johns Hopkins University, Baltimore, Maryland 21218, USA}
\affiliation{Department of Physics and Astronomy, George Mason University, Fairfax, VA 22030, USA}

\author{F.~Tafti}
\affiliation{Department of Physics, Boston College, Chestnut Hill, MA 02467, USA}

\author{J.~Gaudet}
\affiliation{NIST Center for Neutron Research, National Institute of Standards and Technology, Gaithersburg, Maryland 20899, USA}
\affiliation{Department of Materials Science and Eng., University of Maryland, College Park, MD 20742-2115}

\begin{abstract} 
{
We demonstrate that a multi-$\mathbf{k}$ incommensurate magnetic state in the Weyl semimetal CeAlGe gives rise to a singular angular magnetoresistance (SAMR), which is an electrical transport signature capable of detecting magnetic field direction with exceptional precision. In contrast, its sister compound CeAlSi is devoid of both multi-$\mathbf{k}$ order and SAMR. We reveal that both phenomena appear upon 57$\%$ Ge substitution in CeAlSi$_{1-x}$Ge$_x$ and coincide with electronic structure changes that soften the single-ion in-plane anisotropy and enhance Weyl-mediated magnetic interactions. These results unveil a remarkable connection between band topology, electronic transport, and collective magnetism in Weyl semimetals.}
\end{abstract}

\date{\today}

\maketitle
\section{Introduction}

Rare-earth (R) intermetallics \ce{RAlX} (X = Si or Ge) compounds provide a tunable platform to explore the interplay between band topology, magnetism, and electronic transport~\cite{Chang2018}. Their non-centrosymmetric $I4_1\mathrm{md}$ crystal structure (Fig.\ref{Bulk}(a)) stabilizes electronic Weyl nodes that give rise to anomalous transport~\cite{Lyu2020,destraz2020,Yang2020d,yang2021noncollinear,Piva2023a,Piva2023b,yang2023stripe,Dhital2023,alam2023,Kikugawa2024,zhang2024,laha2024,forslund2025}, which are both sensitive to the low-temperature local 4$f$ rare-earth magnetism~\cite{wu2023,lou2023,sanchez2020,sakhya2023,li2023emergence,zhang2023kramers}. In turn, the magnetism can be influenced by conduction electrons via Weyl-mediated Ruderman-Kittel-Kasuya-Yosida (RKKY) interactions, enabling exotic spin textures such as chiral spin density waves or topological magnetic textures~\cite{Puphal2020a,Gaudet2021,yao2023large,drucker2023topology}. Yet, how these intertwined degrees of freedom produce tunable, functional properties remains to be discovered.

In this letter, we present a compelling example by clarifying the origin of the previously reported singular angular magnetoresistance (SAMR) in \ce{CeAlGe}, where a sharp resistivity peak emerges under an in-plane rotating magnetic field in an extremely narrow angular range~\cite{suzuki2019singular}, an attractive feature for potential magnetic sensing applications. It was previously suggested that SAMR arose from the electron scattering on magnetic domain walls (DWs) that form between non-collinear ferromagnetic domains with mismatched nodal Fermi surfaces~\cite{suzuki2019singular}. Here, we demonstrate that this explanation is insufficient and that the emergence of SAMR is linked to an incommensurate magnetic state without which the transport anomaly vanishes entirely. This is demonstrated by chemically tuning the magnetic incommensurability via Ge-to-Si substitution where, in stark contrast with CeAlGe, \ce{CeAlSi_{1-x}Ge_x} with $x<0.57$ exhibits a fully commensurate magnetic order and no SAMR. We further show that the incommensurability in \ce{CeAlGe} emerges from changes in the electronic band structure that weaken the Ce$^{3+}$ single-ion magnetic anisotropy and enhance Weyl-mediated spin-spin interactions.

\section{Results and Analysis}
\subsection{Phase diagram of magnetism and SAMR}

\begin{figure*}[t]
    \includegraphics[width=\textwidth]{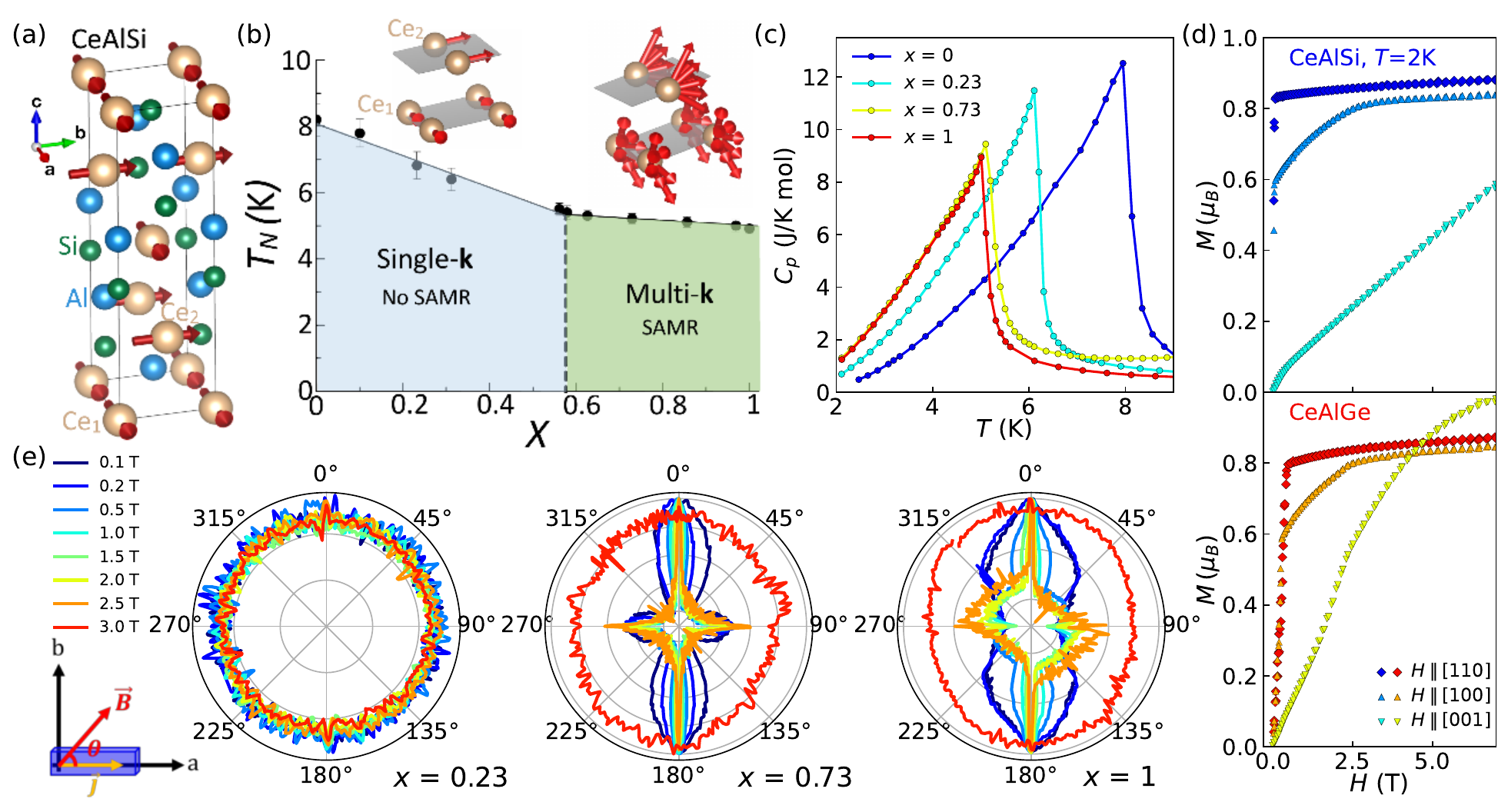}
    \centering
\caption{(a) Nuclear and spin structure of CeAlSi. (b) The magnetic ordering temperature $T_c$ as function of doping $x$ for \ce{CeAlSi_{1-x}Ge_x}. A sketch of the local spin structure is illustrated for the $x<x_c$ and $x>x_c$ phase where $x_c=0.57(4)$. (c) Temperature dependence of $C_p$ for a range of doping $x$. (d) 2~K magnetization data for both CeAlSi (top) and CeAlGe (bottom). (e) Polar plots of the angular dependence of longitudinal resistivity $\rho_{xx}(\theta)$ at several field strengths in three samples with $x=$1, 0.73, and 0.23. Here, $\theta~=~0$ is defined along the $a$-axis and $\rho_{xx}(\theta)$ was normalized to its maximum value for each field.}
    \label{Bulk}
\end{figure*}

We grew single crystals of \ce{CeAlSi_{1-x}Ge_x} with Ge-doping $x$ ranging from 0 to 1. Details of the synthesis procedure and structural characterization are provided in the Supplemental Material (SM). The magnetic and electric phase diagram (Fig.~\ref{Bulk}(b)) of our \ce{CeAlSi_{1-x}Ge_x} crystals were investigated using heat capacity ($C_p$), AC magnetic susceptibility ($\chi$), magnetization ($M$), and longitudinal electrical resistivity ($\rho_{xx}$) measurements. 

Temperature-dependent $C_p$ data, shown in Fig.\ref{Bulk}(c), reveal a single thermodynamic anomaly at low temperatures for all Ge-doped samples, indicating the onset of magnetic ordering. The evolution of the magnetic transition temperature $T_{c}$ as a function of Ge content is presented in Fig.~\ref{Bulk}(b), exhibiting two different slopes across the critical doping $x_c=0.57$. This behavior suggests a possible magnetic phase transition at $x_c=0.57(4)$.

The low-temperature ($T<T_c$) magnetization of CeAlSi and CeAlGe show similar behaviors, with a [110] easy-axis ferromagnetism (Fig.\ref{Bulk}(d)). A difference, however, is observed in the overall spin anisotropy of both compounds where the [001] magnetization curve crosses over the in-plane ones in CeAlGe, but not in CeAlSi. This indicates considerably weaker in-plane anisotropy for CeAlGe.

A striking difference, however, is observed in the angular magneto-resistance properties of $\ce{CeAlSi_{1-x}Ge_x}$. We found that singularities only arise for samples with Ge-doping $x>x_c$ (green phase in Fig.\ref{Bulk}(b)). This is shown in the polar plots of Fig.\ref{Bulk}(e) that present the angular-dependence of the longitudinal in-plane resistivity $\rho_{xx}(\theta)$ for three different dopings ($x~=~0.33,0.73,$ and $1$) at several magnetic fields $H$.

While $\rho_{xx}(\theta)$ at small $x$-values is nearly angle independent at all fields, it acquires a sharp angle dependence in samples with $x>x_c$, which alludes to a magnetic phase transition at $x_c$. Indeed, for $x>x_c$,  the low-$H$ behavior of $\rho_{xx}(\theta)$ starts with a rather smooth angular dependence and no singularities. Upon increasing the field, $\rho_{xx}(\theta)$ shows SAMR for $H$ ranging from $0.5~$T to $2.5~$T. The singularities disappear and a smooth angular dependence of $\rho_{xx}(\theta)$ is recovered for $H~>2.75~$T. Our observed behavior of $\rho_{xx}$ for samples with $x>x_c$ is consistent with a previously reported SAMR in $\ce{CeAlSi_{0.28}Ge_{0.72}}$~\cite{suzuki2019singular}.

\subsection{Onset of incommensurate magnetism at $x_c$}

Motivated by the chemically tunable SAMR in \ce{CeAlSi_{1-x}Ge_x}, we performed small-angle neutron scattering (SANS) on samples with $x~=~0, 0.73,$ and $1$, while we did single-crystal neutron diffraction on samples with $x~=~0, 0.33, 0.73$, and $1$.

\begin{figure}[t]
    \includegraphics[width=\linewidth]{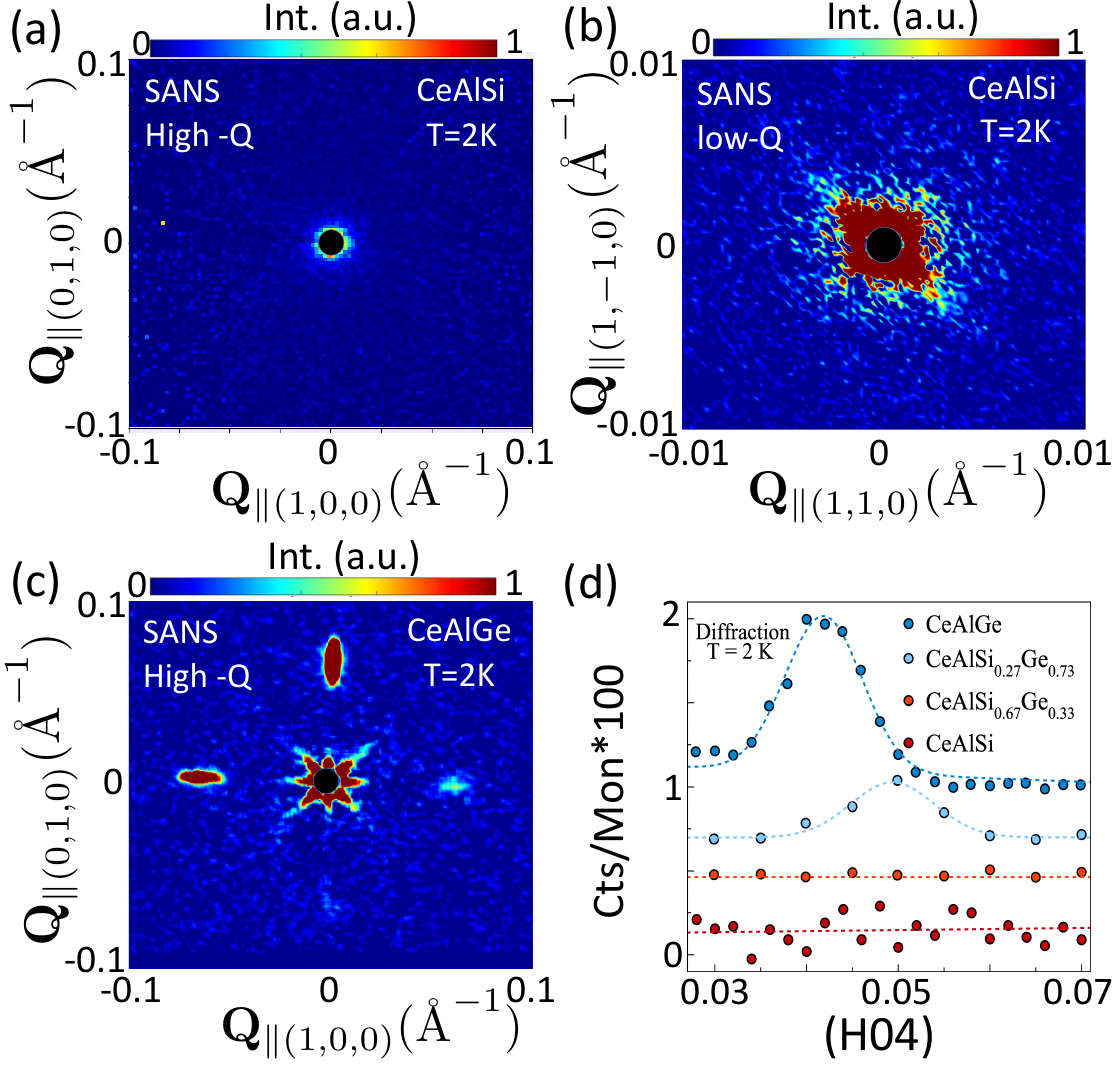}
    \centering
\caption{(a,b) The high-Q (0-0.1~$\angstrom^{-1}$) and low-Q (0-0.01~$\angstrom^{-1}$) SANS data collected for CeAlSi at zero-field. (c) is the high-Q SANS of CeAlGe also collected at zero-field.(d) 2 K single-crystal neutron diffraction of \ce{CeAlSi_{1-x}Ge_x} collected along the (H04) direction of the reciprocal space, which covers the magnetic incommensurability observed only when $x>x_c$.}
    \label{NeutronZF}
\end{figure}

As seen in Fig.\ref{NeutronZF}(a,b), we only detected $\mathbf{Q}~=~\mathbf{0}$ scattering in CeAlSi with SANS, which is better resolved using the low-$\mathbf{Q}$ configuration (Fig.\ref{NeutronZF}(b)). This indicates the spin structure of CeAlSi is fully commensurate with a magnetic ordering vector $\mathbf{k_{com}}=\mathbf{0}$.

Similar to CeAlSi, the SANS of CeAlGe also has $\mathbf{Q}~=~\mathbf{0}$ scattering indicating ferromagnetism. But unlike CeAlSi, the low-temperature high-Q SANS data of CeAlGe (Fig.\ref{NeutronZF}(c)) shows additional Bragg peaks centered at the 4 symmetrically-equivalent reciprocal space positions $\mathbf{k_{inc}}~=~(\pm\eta,0,0)$ and $(0,\pm\eta,0)$ with $\eta=0.042(1)$. We refer to these peaks collectively as $(\eta,0,0)$. These peaks signify $\sim10$~nm modulated spin structures propagating along the $\mathbf{a}$ or $\mathbf{b}$ axes. This spatial modulation decreases with decreasing $x$ to reach $\sim 8.6$~nm at $x~=~0.73$, and finally disappear for $x~<~x_c$.

Like SANS, single-crystal neutron diffraction detects commensurate magnetic Bragg peaks across all Ge doping, but the incommensurate ones are only detected for $x>x_c$ (Fig.\ref{NeutronZF}(d)). Thus, the magnetic ground-state of the $x>x_c$ samples must have both a commensurate $\mathbf{k}_{com}=0$ and an incommensurate $\mathbf{k}_{inc}$ spin component, while it is purely commensurate for $x<x_c$

As previously reported~\cite{suzuki2019singular,yang2021noncollinear}, the commensurate spin component of all doped samples refines best within the spin structure shown in Fig.\ref{Bulk}(a), which corresponds to a non-collinear in-plane ferromagnetic structure where the Ce$_1$ spins point dominantly along the $a$-axis, while the Ce$_2$ spins point dominantly along the $b$-axis (see SM). This commensurate component is stabilized for all doping, so we expect at least four symmetrically-equivalent magnetic domains with total magnetization along the [110], [1-10], [-110], and the [-1-10] direction to be present in all samples. Importantly, DWs are expected to form only for in-plane magnetic fields applied exactly parallel to the $a$ or $b$ axes, i.e. along the directions where SAMR is observed.

The incommensurate component refines best assuming a cycloid forming within the $\mathbf{ac}$ and $\mathbf{bc}$ planes such that the resulting spin structure consists of a non-coplanar conical magnetic texture with an average ferromagnetic background as sketched in Fig.\ref{Bulk}(b) (see SM).

\subsection{Parallel field-dependence of incommensurate magnetism and SAMR}

\begin{figure}[t]
    \includegraphics[width=\linewidth]{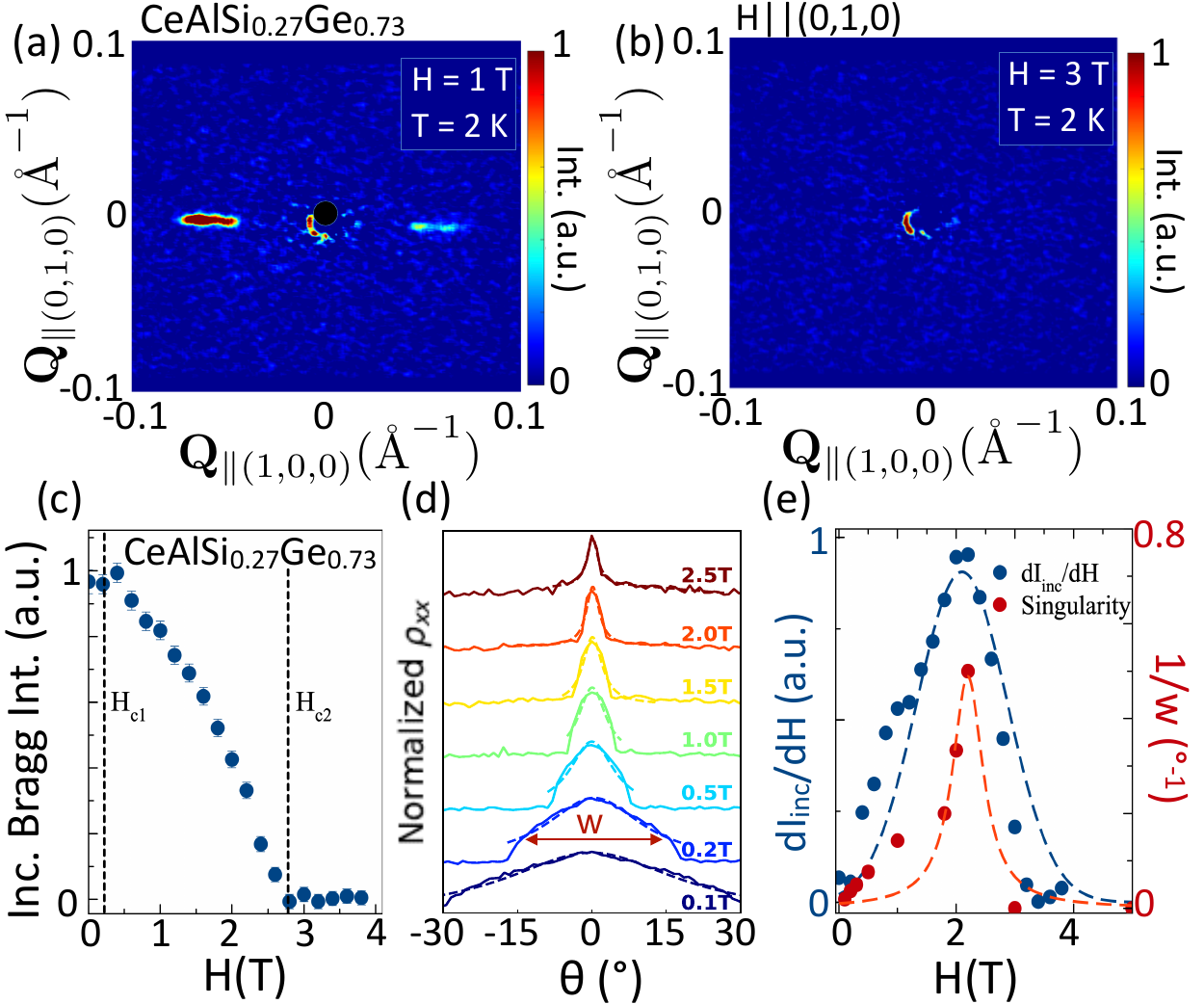}
    \centering
\caption{The 2~K high-Q SANS data for \ce{CeAlSi_{1-x}Ge_x} with $x=0.73$ collected under an in-plane magnetic field H$\parallel$(0,1,0) at (a) $H$~=~1~T and (b) $H$~=~3~T. (c) The field dependence of the incommensurate Bragg peak intensity perpendicular to the field direction. (d) Field-dependence of the angular magnetoresistance of \ce{CeAlSi_{0.27}Ge_{0.73}} near a singularity. Gaussian fits are shown in dashed lines and were used to quantify the singularity, which we defined as 1/W where W is the Gaussian width in $\degree$. (3) Derivative of the field dependence shown in panel (c), overlaid with the inverse width (1/W) of the resisitvity peak of \ce{CeAlSi_{1-x}Ge_x} with $x=0.73$.} 
    \label{NeutronField}
\end{figure}

Given its link to SAMR, we examine the field dependence of the incommensurate order in samples with $x>x_c$, focusing on $x=0.73$, which shows the cleanest SAMR (Fig.\ref{Bulk}(e)). Under a [010] field,  consistent with a cycloidal modulation~\cite{kurumaji2017neel,white2018direct}, the incommensurate modulations perpendicular to the field ($\mathbf{k}_{inc} \perp H$) are selected, reducing the observed diffraction to two of the four $(\eta,0,0)$ peaks for $H_{c1} < H < H_{c2}$ (Fig.\ref{NeutronField}(a)) where $H_{c1}\sim 0.2~$T and $H_{c2}\sim2.75~$T. The two remaining incommensurate Bragg peaks with $\mathbf{k}_{inc} \perp H$ persist at the same ordering vector with similar intensity across $H_{c1}$. This indicates a phase transition from a zero-field 4-$\mathbf{k}_{inc}$ state to a 2-$\mathbf{k}_{inc}$ structure above $H_{c1}$, rather than a reorientation of certain domains, in which case the intensity of the remaining incommensurate peaks would be expected to double at $H_{c1}$. Similar to previous work~\cite{Puphal2020a}, we conclude the zero-field magnetic state of CeAlGe forms with the four $\mathbf{k}~=~(\eta,0,0)$ arms.

As the [010] field further increases, the intensity of the remaining incommensurate Bragg peak gradually decreases and vanishes at $H_{c2}$ (Fig.\ref{NeutronField}(b,c)), signaling a transition from a $2\mathbf{k}_{inc} + \mathbf{k}_{com}$ state to a fully $\mathbf{k}_{com}$ commensurate state. We note that the SAMR is observed for $H_{c1}< H < H_{c2}$ so within the $2\mathbf{k}_{inc} + \mathbf{k}_{com}$ magnetic state. To highlight this connection, the field derivative of the incommensurate peak intensity (Fig.\ref{NeutronField}(c)) and the field-dependent SAMR singularity (Fig.\ref{NeutronField}(d)) are overlaid in Fig.\ref{NeutronField}(e). The latter is characterized by a Gaussian fit to the resistivity peak of the $x=0.73$ sample (Fig.\ref{NeutronField}(d)), where the inverse width ($1/W$) is taken as a phenomenological singularity parameter.

The correlation shown in Fig.~\ref{NeutronField}(e) establishes that the SAMR arises from an in-field incommensurate order. Samples without an incommensurate component do not exhibit SAMR. Thus, we conclude the $2\mathbf{k}_{inc} + \mathbf{k}_{com}$ state, which is not observed along other in-plane field directions, is the proxy for SAMR.

\section{Theory and Discussion}

Our experiments establish a correlation between incommensurate magnetism and SAMR, both appearing at $x>x_c$ and evolving in parallel as a function of magnetic field. We explore three physical mechanisms that work together to stabilize this magnetic order: (1) changes in the crystal electric field (CEF), (2) enhanced spin-orbit coupling (SOC) in Ge relative to Si, and (3)  electronic band structure changes that favor chiral interactions with increasing Ge content. To investigate these possibilities, we perform first-principles calculations for the two end members, CeAlSi and CeAlGe.

\begin{figure}[t]
    \includegraphics[width=\linewidth]{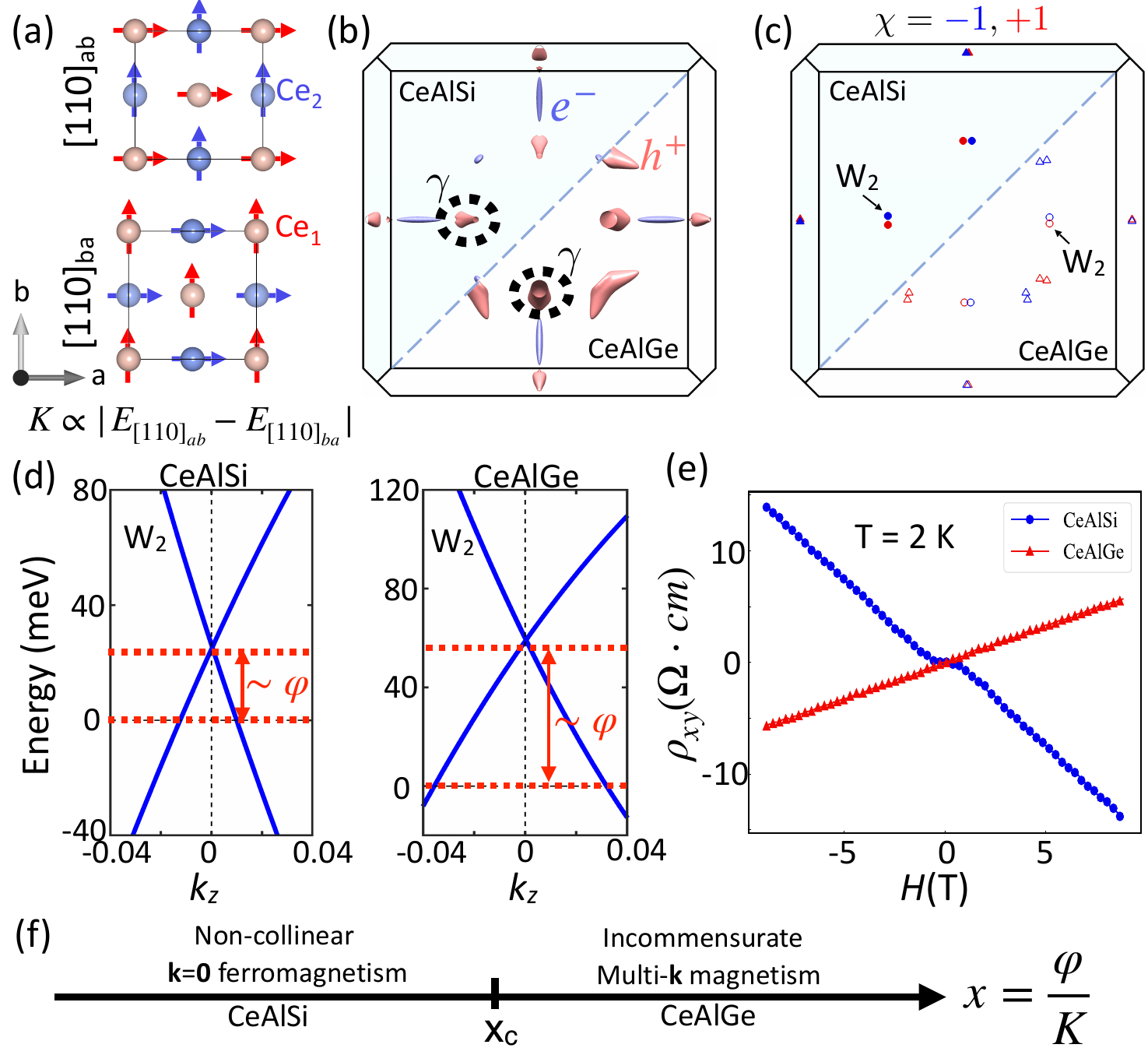}
    \centering
\caption{(a)  Schematic of two non-collinear spin configurations, with Ce$_1$ (red) and Ce$_2$ (blue) moments aligned along orthogonal in-plane directions: Ce$_1$ along $a$ (or $b$) and Ce$_2$ along $b$ (or $a$), denoted as [110]$_{ab}$ and [110]$_{ba}$ respectively. The former is the predicted ground state for CeAlSi. (b) Fermi surface and (c) Weyl node distribution for CeAlSi and CeAlGe in the first Brillouin zone. (d) First-principles calculated $k_z$ dispersion of the W$_2$ nodes for both compounds.  (e)  Hall effect at 2~K showing opposite signs in CeAlSi and CeAlGe. (f) Theoretical magnetic phase diagram as a function of doping $x$, parametrized by the ratio of the multi-spin interactions ($\varphi$) and the CEF anisotropy ($K$). }
    \label{DFT}
\end{figure}

Regarding (1), we find that a stronger CEF at $x<x_c$ favors the commensurate non-collinear ferromagnetic structure (Fig.~\ref{Bulk}(a)). The Ce$_1$ and Ce$_2$ sites share an orthorhombic $2mm$ point group symmetry, giving rise to anisotropic $g$-tensors with $g_a > g_c > g_b$ at Ce$_1$, while the $a$ and $b$ axes are swapped for Ce$_2$ due to the $4_1$ screw symmetry. This naturally promotes non-collinear ordering with each Ce spin aligning along its easy axis. Our DFT calculations reveal that in CeAlSi, the Ce$_1$ ions have their easy and hard axes respectively along the $a$ and $b$ directions, thus favoring the $[110]{ab}$ over the $[110]{ba}$ spin structure (Fig.\ref{DFT}(a)). In contrast, CeAlGe shows little energy difference between these configurations and hosts numerous metastable states. The gradual weakening of the CEF with increasing $x$ is consistent with the reduced in-plane anisotropy of CeAlGe seen in magnetization data (Fig.\ref{Bulk}(d)). 

While the weakening of CEF could precipitate the change of magnetism across $x_c$, it is not enough to induce incommensurability. A potential driver of the incommensurability is the mechanism (2). The spin-orbit coupling is expected to increase with increasing $x$ due to the larger atomic mass of Ge compared to Si. It favors the Dzyaloshinskii-Moriya (DM) interactions known to stabilize incommensurate ordering~\cite{sergienko2006role}.

Finally, (3), the change of band structure between CeAlSi and CeAlGe, can further amplify chiral interactions that promote incommensurability. As shown in Fig.\ref{DFT}(b,c), the Fermi surface of CeAlGe exhibits larger hole pockets compared to CeAlSi. To explain this, the corresponding electronic dispersions of the W$_2$ Weyl nodes are shown in Fig.\ref{DFT}(d). In CeAlGe, the Fermi level lies further from the Weyl nodes, as a result of larger hole pockets. This difference is supported by Hall measurements that show hole-like carriers in CeAlGe and electron-like carriers in CeAlSi (Fig.\ref{DFT}(e)). Notably, theory shows that all chiral interactions, such as DM interaction, induced by Weyl electrons derive their strength from the difference between the Fermi energy ($E_f$) and the node energy \cite{Nikolic2020a,Nikolic2020b}. Thus, chiral interactions are expected to be stronger in CeAlGe with a wider separation between the two energy scales than in CeAlSi (Fig.\ref{DFT}(d)). Among those chiral interactions are multiple-spin interactions of extended range, which are naturally induced by itinerant electrons and hardly intrinsic to well-localized moments. Multiple-spin interactions are known to drive multi-$\mathbf{k_{inc}}$ structures, such as the one observed for CeAlGe~\cite{Ozawa2016}.

While complex magnetic order can often arise from microscopic spin dynamics, the prominence of the Weyl spectrum and the three proposed mechanisms motivate us to explore the potential of Weyl electrons to stabilize the multi-$\mathbf{k}$ order for CeAlGe. Using Monte Carlo, we simulated the spin model
\begin{equation}
    H = -K\sum_{i}({\bf S}_{i}\hat{\bf n}_i)^{2} -\sum_{ij}J_{ij}{\bf S}_{i}{\bf S}_{j} + \sum_{ijk} \Phi_{ijk}\, {\bf S}_i\cdot({\bf S}_j\times{\bf S}_k) \nonumber
\end{equation}
where the easy-axis anisotropy $K$ (mechanism 1; $\hat{\bf n}_i=\hat{\bf x}/\hat{\bf y}$ alternates between even/odd layers) and Heisenberg interaction $J_{ij}$ compete with the Weyl-induced chiral interactions $\Phi_{ijk}$ (mechanisms 2,3). The weyl-mediated DM interaction is omitted because it produces only $2\mathbf{k_{inc}}$ magnetic orders, but 3 and 4-spin interactions $\Phi_{ijk} = c_0 B^{\phantom{x}}_{ijk} + \boldsymbol{\varphi}_{ijk}\cdot\langle{\bf S}\rangle_{ijk}$ that couple the spin chirality to magnetic field ${\bf B}$ and local magnetization $\langle{\bf S}\rangle_{ijk}$ are theoretically predicted \cite{Nikolic2020a,Nikolic2020b} and indeed produce a $4\mathbf{k_{inc}}$ order seen experimentally. Fig.~\ref{DFT}(f) shows the magnetic ground-state selection as a function of the doping parameter $x$ that depends on both the strength of the multi-spin interactions ($\varphi$) and single-ion anisotropy ($K$) (see SM for the model and calculation details). Our model reproduces the doping and temperature evolution of the ground-state between CeAlSi and CeAlGe, confirming the plausibility of the three proposed mechanisms namely CEF, SOC, and band structure effects.

Another salient characteristic of the incommensurate magnetic order in CeAlGe is its role in producing the SAMR. This is different from the previously proposed mechanism for SAMR based on an enhanced scattering from magnetic DWs, which persist only along the four in-plane $\mathbf{a}$-axes~\cite{suzuki2019singular}. While the DW scattering mechanism explains the extreme sensitivity of SAMR to the field angle, since even a slight field misalignment from the $\mathbf{a}$-axis collapses all DWs into a single domain, it stops short of explaining the doping dependence of the effect. Specifically, it does not explain why SAMR is observed only for $x>x_c$.

.

\section{Conclusion}

Our central discovery is that SAMR emerges only at doping above $x_c$ where an incommensurate magnetic component develops. This incommensurate order should modulate the electronic structure~\cite{li2023emergence}, which will certainly boost the Fermi-surface mismatches of adjacent magnetic domains, establishing it as a key ingredient for strong SAMR. Our results also highlights the potential for driving and tuning SAMR in other Weyl semimetals through a combination of CEF, SOC, and band structure effects. These findings thus show how Weyl fermions and collective magnetism can intertwine and give rise to electronic transport behaviors that could be harnessed in the design of future magnetic sensing technologies.

\section*{Acknowledgments}
The work at Boston College (crystal growth as well as transport and magnetization measurements) was funded
by the U.S. Department of Energy, Office of Basic Energy Sciences, Division of Physical Behavior of Materials, under award number DE-SC0023124. This material is based on work supported by the Air Force Office of Scientific Research under award number FA9550-23-1-0124.
The support for neutron scattering was provided by the Center for High-Resolution Neutron Scattering, a partnership between the National Institute of Standards and Technology and the National Science Foundation under Agreement No. DMR-2010792. 
The identification of any commercial product or trade name does not imply endorsement or recommendation by the National Institute of Standards and Technology. 
A portion of this research used resources at the High Flux Isotope Reactor, a DOE Office of Science User Facility operated by the Oak Ridge National Laboratory. The beamtime was allocated to GP-SANS and VERITAS on proposal number IPTS-32074.1. The neutron FONDER experiment was performed under the proposal 24818 at JRR-3, the QUOKKA experiment at ANSTO was performed under proposal P14320, and the SENJU experiment proposal at J-PARC is 2024A0402.
The DFT work at TIFR Mumbai is supported by the Department of Atomic Energy of the Government of India under the Project No. 12-RD-TFR-5.10-0100 and benefited from the computational resources of TIFR Mumbai. The work at Northeastern University was supported by the National Science Foundation through the Expand-QISE award NSF-OMA-2329067 and benefited from the resources of Northeastern University's Advanced Scientific Computation Center, the Discovery Cluster, the Massachusetts Technology Collaborative award MTC-22032, and the Quantum Materials and Sensing Institute.

\bibliography{CeAlGeSi}

\begin{thebibliography}{31}%
\makeatletter
\providecommand \@ifxundefined [1]{%
 \@ifx{#1\undefined}
}%
\providecommand \@ifnum [1]{%
 \ifnum #1\expandafter \@firstoftwo
 \else \expandafter \@secondoftwo
 \fi
}%
\providecommand \@ifx [1]{%
 \ifx #1\expandafter \@firstoftwo
 \else \expandafter \@secondoftwo
 \fi
}%
\providecommand \natexlab [1]{#1}%
\providecommand \enquote  [1]{``#1''}%
\providecommand \bibnamefont  [1]{#1}%
\providecommand \bibfnamefont [1]{#1}%
\providecommand \citenamefont [1]{#1}%
\providecommand \href@noop [0]{\@secondoftwo}%
\providecommand \href [0]{\begingroup \@sanitize@url \@href}%
\providecommand \@href[1]{\@@startlink{#1}\@@href}%
\providecommand \@@href[1]{\endgroup#1\@@endlink}%
\providecommand \@sanitize@url [0]{\catcode `\\12\catcode `\$12\catcode `\&12\catcode `\#12\catcode `\^12\catcode `\_12\catcode `\%12\relax}%
\providecommand \@@startlink[1]{}%
\providecommand \@@endlink[0]{}%
\providecommand \url  [0]{\begingroup\@sanitize@url \@url }%
\providecommand \@url [1]{\endgroup\@href {#1}{\urlprefix }}%
\providecommand \urlprefix  [0]{URL }%
\providecommand \Eprint [0]{\href }%
\providecommand \doibase [0]{http://dx.doi.org/}%
\providecommand \selectlanguage [0]{\@gobble}%
\providecommand \bibinfo  [0]{\@secondoftwo}%
\providecommand \bibfield  [0]{\@secondoftwo}%
\providecommand \translation [1]{[#1]}%
\providecommand \BibitemOpen [0]{}%
\providecommand \bibitemStop [0]{}%
\providecommand \bibitemNoStop [0]{.\EOS\space}%
\providecommand \EOS [0]{\spacefactor3000\relax}%
\providecommand \BibitemShut  [1]{\csname bibitem#1\endcsname}%
\let\auto@bib@innerbib\@empty
%</preamble>
\bibitem [{\citenamefont {Chang}\ \emph {et~al.}(2018)\citenamefont {Chang}, \citenamefont {Singh}, \citenamefont {Xu}, \citenamefont {Bian}, \citenamefont {Huang}, \citenamefont {Hsu}, \citenamefont {Belopolski}, \citenamefont {Alidoust}, \citenamefont {Sanchez}, \citenamefont {Zheng}, \citenamefont {Lu}, \citenamefont {Zhang}, \citenamefont {Bian}, \citenamefont {Chang}, \citenamefont {Jeng}, \citenamefont {Bansil}, \citenamefont {Hsu}, \citenamefont {Jia}, \citenamefont {Neupert}, \citenamefont {Lin},\ and\ \citenamefont {Hasan}}]{Chang2018}%
  \BibitemOpen
  \bibfield  {author} {\bibinfo {author} {\bibfnamefont {G.}~\bibnamefont {Chang}}, \bibinfo {author} {\bibfnamefont {B.}~\bibnamefont {Singh}}, \bibinfo {author} {\bibfnamefont {S.-Y.}\ \bibnamefont {Xu}}, \bibinfo {author} {\bibfnamefont {G.}~\bibnamefont {Bian}}, \bibinfo {author} {\bibfnamefont {S.-M.}\ \bibnamefont {Huang}}, \bibinfo {author} {\bibfnamefont {C.-H.}\ \bibnamefont {Hsu}}, \bibinfo {author} {\bibfnamefont {I.}~\bibnamefont {Belopolski}}, \bibinfo {author} {\bibfnamefont {N.}~\bibnamefont {Alidoust}}, \bibinfo {author} {\bibfnamefont {D.~S.}\ \bibnamefont {Sanchez}}, \bibinfo {author} {\bibfnamefont {H.}~\bibnamefont {Zheng}}, \bibinfo {author} {\bibfnamefont {H.}~\bibnamefont {Lu}}, \bibinfo {author} {\bibfnamefont {X.}~\bibnamefont {Zhang}}, \bibinfo {author} {\bibfnamefont {Y.}~\bibnamefont {Bian}}, \bibinfo {author} {\bibfnamefont {T.-R.}\ \bibnamefont {Chang}}, \bibinfo {author} {\bibfnamefont {H.-T.}\ \bibnamefont {Jeng}}, \bibinfo {author} {\bibfnamefont {A.}~\bibnamefont {Bansil}},
  \bibinfo {author} {\bibfnamefont {H.}~\bibnamefont {Hsu}}, \bibinfo {author} {\bibfnamefont {S.}~\bibnamefont {Jia}}, \bibinfo {author} {\bibfnamefont {T.}~\bibnamefont {Neupert}}, \bibinfo {author} {\bibfnamefont {H.}~\bibnamefont {Lin}}, \ and\ \bibinfo {author} {\bibfnamefont {M.~Z.}\ \bibnamefont {Hasan}},\ }\href {\doibase 10.1103/PhysRevB.97.041104} {\bibfield  {journal} {\bibinfo  {journal} {Phys. Rev. B}\ }\textbf {\bibinfo {volume} {97}},\ \bibinfo {pages} {041104} (\bibinfo {year} {2018})}\BibitemShut {NoStop}%
\bibitem [{\citenamefont {Lyu}\ \emph {et~al.}(2020)\citenamefont {Lyu}, \citenamefont {Xiang}, \citenamefont {Mi}, \citenamefont {Zhao}, \citenamefont {Wang}, \citenamefont {Liu}, \citenamefont {Chen}, \citenamefont {Ren}, \citenamefont {Li},\ and\ \citenamefont {Sun}}]{Lyu2020}%
  \BibitemOpen
  \bibfield  {author} {\bibinfo {author} {\bibfnamefont {M.}~\bibnamefont {Lyu}}, \bibinfo {author} {\bibfnamefont {J.}~\bibnamefont {Xiang}}, \bibinfo {author} {\bibfnamefont {Z.}~\bibnamefont {Mi}}, \bibinfo {author} {\bibfnamefont {H.}~\bibnamefont {Zhao}}, \bibinfo {author} {\bibfnamefont {Z.}~\bibnamefont {Wang}}, \bibinfo {author} {\bibfnamefont {E.}~\bibnamefont {Liu}}, \bibinfo {author} {\bibfnamefont {G.}~\bibnamefont {Chen}}, \bibinfo {author} {\bibfnamefont {Z.}~\bibnamefont {Ren}}, \bibinfo {author} {\bibfnamefont {G.}~\bibnamefont {Li}}, \ and\ \bibinfo {author} {\bibfnamefont {P.}~\bibnamefont {Sun}},\ }\href {\doibase 10.1103/PhysRevB.102.085143} {\bibfield  {journal} {\bibinfo  {journal} {Phys. Rev. B}\ }\textbf {\bibinfo {volume} {102}},\ \bibinfo {pages} {085143} (\bibinfo {year} {2020})}\BibitemShut {NoStop}%
\bibitem [{\citenamefont {Destraz}\ \emph {et~al.}(2020)\citenamefont {Destraz}, \citenamefont {Das}, \citenamefont {Tsirkin}, \citenamefont {Xu}, \citenamefont {Neupert}, \citenamefont {Chang}, \citenamefont {Schilling}, \citenamefont {Grushin}, \citenamefont {Kohlbrecher}, \citenamefont {Keller}, \citenamefont {Puphal}, \citenamefont {Pomjakushina},\ and\ \citenamefont {White}}]{destraz2020}%
  \BibitemOpen
  \bibfield  {author} {\bibinfo {author} {\bibfnamefont {D.}~\bibnamefont {Destraz}}, \bibinfo {author} {\bibfnamefont {L.}~\bibnamefont {Das}}, \bibinfo {author} {\bibfnamefont {S.~S.}\ \bibnamefont {Tsirkin}}, \bibinfo {author} {\bibfnamefont {Y.}~\bibnamefont {Xu}}, \bibinfo {author} {\bibfnamefont {T.}~\bibnamefont {Neupert}}, \bibinfo {author} {\bibfnamefont {J.}~\bibnamefont {Chang}}, \bibinfo {author} {\bibfnamefont {A.}~\bibnamefont {Schilling}}, \bibinfo {author} {\bibfnamefont {A.~G.}\ \bibnamefont {Grushin}}, \bibinfo {author} {\bibfnamefont {J.}~\bibnamefont {Kohlbrecher}}, \bibinfo {author} {\bibfnamefont {L.}~\bibnamefont {Keller}}, \bibinfo {author} {\bibfnamefont {P.}~\bibnamefont {Puphal}}, \bibinfo {author} {\bibfnamefont {E.}~\bibnamefont {Pomjakushina}}, \ and\ \bibinfo {author} {\bibfnamefont {J.~S.}\ \bibnamefont {White}},\ }\href {\doibase 10.1038/s41535-019-0207-7} {\bibfield  {journal} {\bibinfo  {journal} {npj Quantum Mater.}\ }\textbf {\bibinfo {volume} {5}},\ \bibinfo {pages} {5}
  (\bibinfo {year} {2020})}\BibitemShut {NoStop}%
\bibitem [{\citenamefont {Yang}\ \emph {et~al.}(2020)\citenamefont {Yang}, \citenamefont {Singh}, \citenamefont {Lu}, \citenamefont {Huang}, \citenamefont {Bahrami}, \citenamefont {Chiu}, \citenamefont {Graf}, \citenamefont {Huang}, \citenamefont {Wang}, \citenamefont {Lin}, \citenamefont {Torchinsky}, \citenamefont {Bansil},\ and\ \citenamefont {Tafti}}]{Yang2020d}%
  \BibitemOpen
  \bibfield  {author} {\bibinfo {author} {\bibfnamefont {H.-Y.}\ \bibnamefont {Yang}}, \bibinfo {author} {\bibfnamefont {B.}~\bibnamefont {Singh}}, \bibinfo {author} {\bibfnamefont {B.}~\bibnamefont {Lu}}, \bibinfo {author} {\bibfnamefont {C.-Y.}\ \bibnamefont {Huang}}, \bibinfo {author} {\bibfnamefont {F.}~\bibnamefont {Bahrami}}, \bibinfo {author} {\bibfnamefont {W.-C.}\ \bibnamefont {Chiu}}, \bibinfo {author} {\bibfnamefont {D.}~\bibnamefont {Graf}}, \bibinfo {author} {\bibfnamefont {S.-M.}\ \bibnamefont {Huang}}, \bibinfo {author} {\bibfnamefont {B.}~\bibnamefont {Wang}}, \bibinfo {author} {\bibfnamefont {H.}~\bibnamefont {Lin}}, \bibinfo {author} {\bibfnamefont {D.}~\bibnamefont {Torchinsky}}, \bibinfo {author} {\bibfnamefont {A.}~\bibnamefont {Bansil}}, \ and\ \bibinfo {author} {\bibfnamefont {F.}~\bibnamefont {Tafti}},\ }\href {\doibase 10.1063/1.5132958} {\bibfield  {journal} {\bibinfo  {journal} {APL Mater.}\ }\textbf {\bibinfo {volume} {8}},\ \bibinfo {pages} {011111} (\bibinfo {year}
  {2020})}\BibitemShut {NoStop}%
\bibitem [{\citenamefont {Yang}\ \emph {et~al.}(2021)\citenamefont {Yang}, \citenamefont {Singh}, \citenamefont {Gaudet}, \citenamefont {Lu}, \citenamefont {Huang}, \citenamefont {Chiu}, \citenamefont {Huang}, \citenamefont {Wang}, \citenamefont {Bahrami}, \citenamefont {Xu} \emph {et~al.}}]{yang2021noncollinear}%
  \BibitemOpen
  \bibfield  {author} {\bibinfo {author} {\bibfnamefont {H.-Y.}\ \bibnamefont {Yang}}, \bibinfo {author} {\bibfnamefont {B.}~\bibnamefont {Singh}}, \bibinfo {author} {\bibfnamefont {J.}~\bibnamefont {Gaudet}}, \bibinfo {author} {\bibfnamefont {B.}~\bibnamefont {Lu}}, \bibinfo {author} {\bibfnamefont {C.-Y.}\ \bibnamefont {Huang}}, \bibinfo {author} {\bibfnamefont {W.-C.}\ \bibnamefont {Chiu}}, \bibinfo {author} {\bibfnamefont {S.-M.}\ \bibnamefont {Huang}}, \bibinfo {author} {\bibfnamefont {B.}~\bibnamefont {Wang}}, \bibinfo {author} {\bibfnamefont {F.}~\bibnamefont {Bahrami}}, \bibinfo {author} {\bibfnamefont {B.}~\bibnamefont {Xu}},  \emph {et~al.},\ }\href {\doibase https://doi.org/10.1103/PhysRevB.103.115143} {\bibfield  {journal} {\bibinfo  {journal} {Phys. Rev. B}\ }\textbf {\bibinfo {volume} {103}},\ \bibinfo {pages} {115143} (\bibinfo {year} {2021})}\BibitemShut {NoStop}%
\bibitem [{\citenamefont {Piva}\ \emph {et~al.}(2023{\natexlab{a}})\citenamefont {Piva}, \citenamefont {Souza}, \citenamefont {Lombardi}, \citenamefont {Pakuszewski}, \citenamefont {Adriano}, \citenamefont {Pagliuso},\ and\ \citenamefont {Nicklas}}]{Piva2023a}%
  \BibitemOpen
  \bibfield  {author} {\bibinfo {author} {\bibfnamefont {M.~M.}\ \bibnamefont {Piva}}, \bibinfo {author} {\bibfnamefont {J.~C.}\ \bibnamefont {Souza}}, \bibinfo {author} {\bibfnamefont {G.~A.}\ \bibnamefont {Lombardi}}, \bibinfo {author} {\bibfnamefont {K.~R.}\ \bibnamefont {Pakuszewski}}, \bibinfo {author} {\bibfnamefont {C.}~\bibnamefont {Adriano}}, \bibinfo {author} {\bibfnamefont {P.~G.}\ \bibnamefont {Pagliuso}}, \ and\ \bibinfo {author} {\bibfnamefont {M.}~\bibnamefont {Nicklas}},\ }\href {\doibase 10.1103/PhysRevMaterials.7.074204} {\bibfield  {journal} {\bibinfo  {journal} {Phys. Rev. Mater.}\ }\textbf {\bibinfo {volume} {7}},\ \bibinfo {pages} {074204} (\bibinfo {year} {2023}{\natexlab{a}})}\BibitemShut {NoStop}%
\bibitem [{\citenamefont {Piva}\ \emph {et~al.}(2023{\natexlab{b}})\citenamefont {Piva}, \citenamefont {Souza}, \citenamefont {Brousseau-Couture}, \citenamefont {Sorn}, \citenamefont {Pakuszewski}, \citenamefont {John}, \citenamefont {Adriano}, \citenamefont {C{\^{o}}t{\'{e}}}, \citenamefont {Pagliuso}, \citenamefont {Paramekanti},\ and\ \citenamefont {Nicklas}}]{Piva2023b}%
  \BibitemOpen
  \bibfield  {author} {\bibinfo {author} {\bibfnamefont {M.~M.}\ \bibnamefont {Piva}}, \bibinfo {author} {\bibfnamefont {J.~C.}\ \bibnamefont {Souza}}, \bibinfo {author} {\bibfnamefont {V.}~\bibnamefont {Brousseau-Couture}}, \bibinfo {author} {\bibfnamefont {S.}~\bibnamefont {Sorn}}, \bibinfo {author} {\bibfnamefont {K.~R.}\ \bibnamefont {Pakuszewski}}, \bibinfo {author} {\bibfnamefont {J.~K.}\ \bibnamefont {John}}, \bibinfo {author} {\bibfnamefont {C.}~\bibnamefont {Adriano}}, \bibinfo {author} {\bibfnamefont {M.}~\bibnamefont {C{\^{o}}t{\'{e}}}}, \bibinfo {author} {\bibfnamefont {P.~G.}\ \bibnamefont {Pagliuso}}, \bibinfo {author} {\bibfnamefont {A.}~\bibnamefont {Paramekanti}}, \ and\ \bibinfo {author} {\bibfnamefont {M.}~\bibnamefont {Nicklas}},\ }\href {\doibase 10.1103/PhysRevResearch.5.013068} {\bibfield  {journal} {\bibinfo  {journal} {Phys. Rev. Res.}\ }\textbf {\bibinfo {volume} {5}},\ \bibinfo {pages} {013068} (\bibinfo {year} {2023}{\natexlab{b}})}\BibitemShut {NoStop}%
\bibitem [{\citenamefont {Yang}\ \emph {et~al.}(2023)\citenamefont {Yang}, \citenamefont {Gaudet}, \citenamefont {Verma}, \citenamefont {Baidya}, \citenamefont {Bahrami}, \citenamefont {Yao}, \citenamefont {Huang}, \citenamefont {DeBeer-Schmitt}, \citenamefont {Aczel}, \citenamefont {Xu} \emph {et~al.}}]{yang2023stripe}%
  \BibitemOpen
  \bibfield  {author} {\bibinfo {author} {\bibfnamefont {H.-Y.}\ \bibnamefont {Yang}}, \bibinfo {author} {\bibfnamefont {J.}~\bibnamefont {Gaudet}}, \bibinfo {author} {\bibfnamefont {R.}~\bibnamefont {Verma}}, \bibinfo {author} {\bibfnamefont {S.}~\bibnamefont {Baidya}}, \bibinfo {author} {\bibfnamefont {F.}~\bibnamefont {Bahrami}}, \bibinfo {author} {\bibfnamefont {X.}~\bibnamefont {Yao}}, \bibinfo {author} {\bibfnamefont {C.-Y.}\ \bibnamefont {Huang}}, \bibinfo {author} {\bibfnamefont {L.}~\bibnamefont {DeBeer-Schmitt}}, \bibinfo {author} {\bibfnamefont {A.~A.}\ \bibnamefont {Aczel}}, \bibinfo {author} {\bibfnamefont {G.}~\bibnamefont {Xu}},  \emph {et~al.},\ }\href {\doibase https://doi.org/10.1103/PhysRevMaterials.7.034202} {\bibfield  {journal} {\bibinfo  {journal} {Phys. Rev. Mat.}\ }\textbf {\bibinfo {volume} {7}},\ \bibinfo {pages} {034202} (\bibinfo {year} {2023})}\BibitemShut {NoStop}%
\bibitem [{\citenamefont {Dhital}\ \emph {et~al.}(2023)\citenamefont {Dhital}, \citenamefont {Dally}, \citenamefont {Ruvalcaba}, \citenamefont {Gonzalez-Hernandez}, \citenamefont {Guerrero-Sanchez}, \citenamefont {Cao}, \citenamefont {Zhang}, \citenamefont {Tian}, \citenamefont {Wu}, \citenamefont {Frontzek}, \citenamefont {Karna}, \citenamefont {Meads}, \citenamefont {Wilson}, \citenamefont {Chapai}, \citenamefont {Graf}, \citenamefont {Bacsa}, \citenamefont {Jin},\ and\ \citenamefont {DiTusa}}]{Dhital2023}%
  \BibitemOpen
  \bibfield  {author} {\bibinfo {author} {\bibfnamefont {C.}~\bibnamefont {Dhital}}, \bibinfo {author} {\bibfnamefont {R.~L.}\ \bibnamefont {Dally}}, \bibinfo {author} {\bibfnamefont {R.}~\bibnamefont {Ruvalcaba}}, \bibinfo {author} {\bibfnamefont {R.}~\bibnamefont {Gonzalez-Hernandez}}, \bibinfo {author} {\bibfnamefont {J.}~\bibnamefont {Guerrero-Sanchez}}, \bibinfo {author} {\bibfnamefont {H.~B.}\ \bibnamefont {Cao}}, \bibinfo {author} {\bibfnamefont {Q.}~\bibnamefont {Zhang}}, \bibinfo {author} {\bibfnamefont {W.}~\bibnamefont {Tian}}, \bibinfo {author} {\bibfnamefont {Y.}~\bibnamefont {Wu}}, \bibinfo {author} {\bibfnamefont {M.~D.}\ \bibnamefont {Frontzek}}, \bibinfo {author} {\bibfnamefont {S.~K.}\ \bibnamefont {Karna}}, \bibinfo {author} {\bibfnamefont {A.}~\bibnamefont {Meads}}, \bibinfo {author} {\bibfnamefont {B.}~\bibnamefont {Wilson}}, \bibinfo {author} {\bibfnamefont {R.}~\bibnamefont {Chapai}}, \bibinfo {author} {\bibfnamefont {D.}~\bibnamefont {Graf}}, \bibinfo {author} {\bibfnamefont
  {J.}~\bibnamefont {Bacsa}}, \bibinfo {author} {\bibfnamefont {R.}~\bibnamefont {Jin}}, \ and\ \bibinfo {author} {\bibfnamefont {J.~F.}\ \bibnamefont {DiTusa}},\ }\href {\doibase 10.1103/PhysRevB.107.224414} {\bibfield  {journal} {\bibinfo  {journal} {Phys. Rev. B}\ }\textbf {\bibinfo {volume} {107}},\ \bibinfo {pages} {224414} (\bibinfo {year} {2023})}\BibitemShut {NoStop}%
\bibitem [{\citenamefont {Alam}\ \emph {et~al.}(2023)\citenamefont {Alam}, \citenamefont {Fakhredine}, \citenamefont {Ahmad}, \citenamefont {Tanwar}, \citenamefont {Yang}, \citenamefont {Tafti}, \citenamefont {Cuono}, \citenamefont {Islam}, \citenamefont {Singh}, \citenamefont {Lynnyk} \emph {et~al.}}]{alam2023}%
  \BibitemOpen
  \bibfield  {author} {\bibinfo {author} {\bibfnamefont {M.~S.}\ \bibnamefont {Alam}}, \bibinfo {author} {\bibfnamefont {A.}~\bibnamefont {Fakhredine}}, \bibinfo {author} {\bibfnamefont {M.}~\bibnamefont {Ahmad}}, \bibinfo {author} {\bibfnamefont {P.}~\bibnamefont {Tanwar}}, \bibinfo {author} {\bibfnamefont {H.-Y.}\ \bibnamefont {Yang}}, \bibinfo {author} {\bibfnamefont {F.}~\bibnamefont {Tafti}}, \bibinfo {author} {\bibfnamefont {G.}~\bibnamefont {Cuono}}, \bibinfo {author} {\bibfnamefont {R.}~\bibnamefont {Islam}}, \bibinfo {author} {\bibfnamefont {B.}~\bibnamefont {Singh}}, \bibinfo {author} {\bibfnamefont {A.}~\bibnamefont {Lynnyk}},  \emph {et~al.},\ }\href {\doibase https://doi.org/10.1103/PhysRevB.107.085102} {\bibfield  {journal} {\bibinfo  {journal} {Phys. Rev. B}\ }\textbf {\bibinfo {volume} {107}},\ \bibinfo {pages} {085102} (\bibinfo {year} {2023})}\BibitemShut {NoStop}%
\bibitem [{\citenamefont {Kikugawa}\ \emph {et~al.}(2024)\citenamefont {Kikugawa}, \citenamefont {Uji},\ and\ \citenamefont {Terashima}}]{Kikugawa2024}%
  \BibitemOpen
  \bibfield  {author} {\bibinfo {author} {\bibfnamefont {N.}~\bibnamefont {Kikugawa}}, \bibinfo {author} {\bibfnamefont {S.}~\bibnamefont {Uji}}, \ and\ \bibinfo {author} {\bibfnamefont {T.}~\bibnamefont {Terashima}},\ }\href {\doibase 10.1103/PhysRevB.109.035143} {\bibfield  {journal} {\bibinfo  {journal} {Phys. Rev. B}\ }\textbf {\bibinfo {volume} {109}},\ \bibinfo {pages} {035143} (\bibinfo {year} {2024})}\BibitemShut {NoStop}%
\bibitem [{\citenamefont {Zhang}\ \emph {et~al.}(2024)\citenamefont {Zhang}, \citenamefont {Tu}, \citenamefont {Li}, \citenamefont {Tang}, \citenamefont {Nie}, \citenamefont {Li}, \citenamefont {Li}, \citenamefont {Qi}, \citenamefont {Wu}, \citenamefont {Zhou} \emph {et~al.}}]{zhang2024}%
  \BibitemOpen
  \bibfield  {author} {\bibinfo {author} {\bibfnamefont {N.}~\bibnamefont {Zhang}}, \bibinfo {author} {\bibfnamefont {D.}~\bibnamefont {Tu}}, \bibinfo {author} {\bibfnamefont {D.}~\bibnamefont {Li}}, \bibinfo {author} {\bibfnamefont {K.}~\bibnamefont {Tang}}, \bibinfo {author} {\bibfnamefont {L.}~\bibnamefont {Nie}}, \bibinfo {author} {\bibfnamefont {H.}~\bibnamefont {Li}}, \bibinfo {author} {\bibfnamefont {H.}~\bibnamefont {Li}}, \bibinfo {author} {\bibfnamefont {T.}~\bibnamefont {Qi}}, \bibinfo {author} {\bibfnamefont {T.}~\bibnamefont {Wu}}, \bibinfo {author} {\bibfnamefont {J.}~\bibnamefont {Zhou}},  \emph {et~al.},\ }\href {\doibase https://doi.org/10.1038/s41467-024-54632-0} {\bibfield  {journal} {\bibinfo  {journal} {Nat. Commun.}\ }\textbf {\bibinfo {volume} {15}},\ \bibinfo {pages} {10255} (\bibinfo {year} {2024})}\BibitemShut {NoStop}%
\bibitem [{\citenamefont {Laha}\ \emph {et~al.}(2024)\citenamefont {Laha}, \citenamefont {Kundu}, \citenamefont {Aryal}, \citenamefont {Bozin}, \citenamefont {Yao}, \citenamefont {Paone}, \citenamefont {Rajapitamahuni}, \citenamefont {Vescovo}, \citenamefont {Valla}, \citenamefont {Abeykoon} \emph {et~al.}}]{laha2024}%
  \BibitemOpen
  \bibfield  {author} {\bibinfo {author} {\bibfnamefont {A.}~\bibnamefont {Laha}}, \bibinfo {author} {\bibfnamefont {A.~K.}\ \bibnamefont {Kundu}}, \bibinfo {author} {\bibfnamefont {N.}~\bibnamefont {Aryal}}, \bibinfo {author} {\bibfnamefont {E.~S.}\ \bibnamefont {Bozin}}, \bibinfo {author} {\bibfnamefont {J.}~\bibnamefont {Yao}}, \bibinfo {author} {\bibfnamefont {S.}~\bibnamefont {Paone}}, \bibinfo {author} {\bibfnamefont {A.}~\bibnamefont {Rajapitamahuni}}, \bibinfo {author} {\bibfnamefont {E.}~\bibnamefont {Vescovo}}, \bibinfo {author} {\bibfnamefont {T.}~\bibnamefont {Valla}}, \bibinfo {author} {\bibfnamefont {M.}~\bibnamefont {Abeykoon}},  \emph {et~al.},\ }\href {\doibase https://doi.org/10.1103/PhysRevB.109.035120} {\bibfield  {journal} {\bibinfo  {journal} {Phys. Rev. B}\ }\textbf {\bibinfo {volume} {109}},\ \bibinfo {pages} {035120} (\bibinfo {year} {2024})}\BibitemShut {NoStop}%
\bibitem [{\citenamefont {Forslund}\ \emph {et~al.}(2025)\citenamefont {Forslund}, \citenamefont {Liu}, \citenamefont {Shin}, \citenamefont {Lin}, \citenamefont {Horio}, \citenamefont {Wang}, \citenamefont {Kramer}, \citenamefont {Mukherjee}, \citenamefont {Kim}, \citenamefont {Cacho} \emph {et~al.}}]{forslund2025}%
  \BibitemOpen
  \bibfield  {author} {\bibinfo {author} {\bibfnamefont {O.~K.}\ \bibnamefont {Forslund}}, \bibinfo {author} {\bibfnamefont {X.}~\bibnamefont {Liu}}, \bibinfo {author} {\bibfnamefont {S.}~\bibnamefont {Shin}}, \bibinfo {author} {\bibfnamefont {C.}~\bibnamefont {Lin}}, \bibinfo {author} {\bibfnamefont {M.}~\bibnamefont {Horio}}, \bibinfo {author} {\bibfnamefont {Q.}~\bibnamefont {Wang}}, \bibinfo {author} {\bibfnamefont {K.}~\bibnamefont {Kramer}}, \bibinfo {author} {\bibfnamefont {S.}~\bibnamefont {Mukherjee}}, \bibinfo {author} {\bibfnamefont {T.}~\bibnamefont {Kim}}, \bibinfo {author} {\bibfnamefont {C.}~\bibnamefont {Cacho}},  \emph {et~al.},\ }\href {\doibase https://doi.org/10.1103/PhysRevLett.134.126602} {\bibfield  {journal} {\bibinfo  {journal} {Phys. Rev. Lett.}\ }\textbf {\bibinfo {volume} {134}},\ \bibinfo {pages} {126602} (\bibinfo {year} {2025})}\BibitemShut {NoStop}%
\bibitem [{\citenamefont {Wu}\ \emph {et~al.}(2023)\citenamefont {Wu}, \citenamefont {Chi}, \citenamefont {Zuo}, \citenamefont {Xu}, \citenamefont {Zhao}, \citenamefont {Luo},\ and\ \citenamefont {Zhu}}]{wu2023}%
  \BibitemOpen
  \bibfield  {author} {\bibinfo {author} {\bibfnamefont {L.}~\bibnamefont {Wu}}, \bibinfo {author} {\bibfnamefont {S.}~\bibnamefont {Chi}}, \bibinfo {author} {\bibfnamefont {H.}~\bibnamefont {Zuo}}, \bibinfo {author} {\bibfnamefont {G.}~\bibnamefont {Xu}}, \bibinfo {author} {\bibfnamefont {L.}~\bibnamefont {Zhao}}, \bibinfo {author} {\bibfnamefont {Y.}~\bibnamefont {Luo}}, \ and\ \bibinfo {author} {\bibfnamefont {Z.}~\bibnamefont {Zhu}},\ }\href {\doibase https://doi.org/10.1038/s41535-023-00537-y} {\bibfield  {journal} {\bibinfo  {journal} {npj Quantum Mater.}\ }\textbf {\bibinfo {volume} {8}},\ \bibinfo {pages} {4} (\bibinfo {year} {2023})}\BibitemShut {NoStop}%
\bibitem [{\citenamefont {Lou}\ \emph {et~al.}(2023)\citenamefont {Lou}, \citenamefont {Fedorov}, \citenamefont {Zhao}, \citenamefont {Yaresko}, \citenamefont {B{\"u}chner},\ and\ \citenamefont {Borisenko}}]{lou2023}%
  \BibitemOpen
  \bibfield  {author} {\bibinfo {author} {\bibfnamefont {R.}~\bibnamefont {Lou}}, \bibinfo {author} {\bibfnamefont {A.}~\bibnamefont {Fedorov}}, \bibinfo {author} {\bibfnamefont {L.}~\bibnamefont {Zhao}}, \bibinfo {author} {\bibfnamefont {A.}~\bibnamefont {Yaresko}}, \bibinfo {author} {\bibfnamefont {B.}~\bibnamefont {B{\"u}chner}}, \ and\ \bibinfo {author} {\bibfnamefont {S.}~\bibnamefont {Borisenko}},\ }\href {\doibase https://doi.org/10.1103/PhysRevB.107.035158} {\bibfield  {journal} {\bibinfo  {journal} {Phys. Rev. B}\ }\textbf {\bibinfo {volume} {107}},\ \bibinfo {pages} {035158} (\bibinfo {year} {2023})}\BibitemShut {NoStop}%
\bibitem [{\citenamefont {Sanchez}\ \emph {et~al.}(2020)\citenamefont {Sanchez}, \citenamefont {Chang}, \citenamefont {Belopolski}, \citenamefont {Lu}, \citenamefont {Yin}, \citenamefont {Alidoust}, \citenamefont {Xu}, \citenamefont {Cochran}, \citenamefont {Zhang}, \citenamefont {Bian} \emph {et~al.}}]{sanchez2020}%
  \BibitemOpen
  \bibfield  {author} {\bibinfo {author} {\bibfnamefont {D.~S.}\ \bibnamefont {Sanchez}}, \bibinfo {author} {\bibfnamefont {G.}~\bibnamefont {Chang}}, \bibinfo {author} {\bibfnamefont {I.}~\bibnamefont {Belopolski}}, \bibinfo {author} {\bibfnamefont {H.}~\bibnamefont {Lu}}, \bibinfo {author} {\bibfnamefont {J.-X.}\ \bibnamefont {Yin}}, \bibinfo {author} {\bibfnamefont {N.}~\bibnamefont {Alidoust}}, \bibinfo {author} {\bibfnamefont {X.}~\bibnamefont {Xu}}, \bibinfo {author} {\bibfnamefont {T.~A.}\ \bibnamefont {Cochran}}, \bibinfo {author} {\bibfnamefont {X.}~\bibnamefont {Zhang}}, \bibinfo {author} {\bibfnamefont {Y.}~\bibnamefont {Bian}},  \emph {et~al.},\ }\href {\doibase https://doi.org/10.1038/s41467-020-16879-1} {\bibfield  {journal} {\bibinfo  {journal} {Nat. Commun.}\ }\textbf {\bibinfo {volume} {11}},\ \bibinfo {pages} {3356} (\bibinfo {year} {2020})}\BibitemShut {NoStop}%
\bibitem [{\citenamefont {Sakhya}\ \emph {et~al.}(2023)\citenamefont {Sakhya}, \citenamefont {Huang}, \citenamefont {Dhakal}, \citenamefont {Gao}, \citenamefont {Regmi}, \citenamefont {Wang}, \citenamefont {Wen}, \citenamefont {He}, \citenamefont {Yao}, \citenamefont {Smith} \emph {et~al.}}]{sakhya2023}%
  \BibitemOpen
  \bibfield  {author} {\bibinfo {author} {\bibfnamefont {A.~P.}\ \bibnamefont {Sakhya}}, \bibinfo {author} {\bibfnamefont {C.-Y.}\ \bibnamefont {Huang}}, \bibinfo {author} {\bibfnamefont {G.}~\bibnamefont {Dhakal}}, \bibinfo {author} {\bibfnamefont {X.-J.}\ \bibnamefont {Gao}}, \bibinfo {author} {\bibfnamefont {S.}~\bibnamefont {Regmi}}, \bibinfo {author} {\bibfnamefont {B.}~\bibnamefont {Wang}}, \bibinfo {author} {\bibfnamefont {W.}~\bibnamefont {Wen}}, \bibinfo {author} {\bibfnamefont {R.-H.}\ \bibnamefont {He}}, \bibinfo {author} {\bibfnamefont {X.}~\bibnamefont {Yao}}, \bibinfo {author} {\bibfnamefont {R.}~\bibnamefont {Smith}},  \emph {et~al.},\ }\href {\doibase https://doi.org/10.1103/PhysRevMaterials.7.L051202} {\bibfield  {journal} {\bibinfo  {journal} {Phys. Rev. Mat.}\ }\textbf {\bibinfo {volume} {7}},\ \bibinfo {pages} {L051202} (\bibinfo {year} {2023})}\BibitemShut {NoStop}%
\bibitem [{\citenamefont {Li}\ \emph {et~al.}(2023)\citenamefont {Li}, \citenamefont {Zhang}, \citenamefont {Wang}, \citenamefont {Liu}, \citenamefont {Guo}, \citenamefont {Rienks}, \citenamefont {Chen}, \citenamefont {Bertran}, \citenamefont {Yang}, \citenamefont {Phuyal} \emph {et~al.}}]{li2023emergence}%
  \BibitemOpen
  \bibfield  {author} {\bibinfo {author} {\bibfnamefont {C.}~\bibnamefont {Li}}, \bibinfo {author} {\bibfnamefont {J.}~\bibnamefont {Zhang}}, \bibinfo {author} {\bibfnamefont {Y.}~\bibnamefont {Wang}}, \bibinfo {author} {\bibfnamefont {H.}~\bibnamefont {Liu}}, \bibinfo {author} {\bibfnamefont {Q.}~\bibnamefont {Guo}}, \bibinfo {author} {\bibfnamefont {E.}~\bibnamefont {Rienks}}, \bibinfo {author} {\bibfnamefont {W.}~\bibnamefont {Chen}}, \bibinfo {author} {\bibfnamefont {F.}~\bibnamefont {Bertran}}, \bibinfo {author} {\bibfnamefont {H.}~\bibnamefont {Yang}}, \bibinfo {author} {\bibfnamefont {D.}~\bibnamefont {Phuyal}},  \emph {et~al.},\ }\href {\doibase https://doi.org/10.1038/s41467-023-42996-8} {\bibfield  {journal} {\bibinfo  {journal} {Nat. Commun.}\ }\textbf {\bibinfo {volume} {14}},\ \bibinfo {pages} {7185} (\bibinfo {year} {2023})}\BibitemShut {NoStop}%
\bibitem [{\citenamefont {Zhang}\ \emph {et~al.}(2023)\citenamefont {Zhang}, \citenamefont {Gao}, \citenamefont {Gao}, \citenamefont {Lei}, \citenamefont {Ni}, \citenamefont {Oh}, \citenamefont {Huang}, \citenamefont {Yue}, \citenamefont {Zonno}, \citenamefont {Gorovikov} \emph {et~al.}}]{zhang2023kramers}%
  \BibitemOpen
  \bibfield  {author} {\bibinfo {author} {\bibfnamefont {Y.}~\bibnamefont {Zhang}}, \bibinfo {author} {\bibfnamefont {Y.}~\bibnamefont {Gao}}, \bibinfo {author} {\bibfnamefont {X.-J.}\ \bibnamefont {Gao}}, \bibinfo {author} {\bibfnamefont {S.}~\bibnamefont {Lei}}, \bibinfo {author} {\bibfnamefont {Z.}~\bibnamefont {Ni}}, \bibinfo {author} {\bibfnamefont {J.~S.}\ \bibnamefont {Oh}}, \bibinfo {author} {\bibfnamefont {J.}~\bibnamefont {Huang}}, \bibinfo {author} {\bibfnamefont {Z.}~\bibnamefont {Yue}}, \bibinfo {author} {\bibfnamefont {M.}~\bibnamefont {Zonno}}, \bibinfo {author} {\bibfnamefont {S.}~\bibnamefont {Gorovikov}},  \emph {et~al.},\ }\href {\doibase https://doi.org/10.1038/s42005-023-01257-2} {\bibfield  {journal} {\bibinfo  {journal} {Commun. Phys.}\ }\textbf {\bibinfo {volume} {6}},\ \bibinfo {pages} {134} (\bibinfo {year} {2023})}\BibitemShut {NoStop}%
\bibitem [{\citenamefont {Puphal}\ \emph {et~al.}(2020)\citenamefont {Puphal}, \citenamefont {Pomjakushin}, \citenamefont {Kanazawa}, \citenamefont {Ukleev}, \citenamefont {Gawryluk}, \citenamefont {Ma}, \citenamefont {Naamneh}, \citenamefont {Plumb}, \citenamefont {Keller}, \citenamefont {Cubitt}, \citenamefont {Pomjakushina},\ and\ \citenamefont {White}}]{Puphal2020a}%
  \BibitemOpen
  \bibfield  {author} {\bibinfo {author} {\bibfnamefont {P.}~\bibnamefont {Puphal}}, \bibinfo {author} {\bibfnamefont {V.}~\bibnamefont {Pomjakushin}}, \bibinfo {author} {\bibfnamefont {N.}~\bibnamefont {Kanazawa}}, \bibinfo {author} {\bibfnamefont {V.}~\bibnamefont {Ukleev}}, \bibinfo {author} {\bibfnamefont {D.~J.}\ \bibnamefont {Gawryluk}}, \bibinfo {author} {\bibfnamefont {J.}~\bibnamefont {Ma}}, \bibinfo {author} {\bibfnamefont {M.}~\bibnamefont {Naamneh}}, \bibinfo {author} {\bibfnamefont {N.~C.}\ \bibnamefont {Plumb}}, \bibinfo {author} {\bibfnamefont {L.}~\bibnamefont {Keller}}, \bibinfo {author} {\bibfnamefont {R.}~\bibnamefont {Cubitt}}, \bibinfo {author} {\bibfnamefont {E.}~\bibnamefont {Pomjakushina}}, \ and\ \bibinfo {author} {\bibfnamefont {J.~S.}\ \bibnamefont {White}},\ }\href {\doibase 10.1103/PhysRevLett.124.017202} {\bibfield  {journal} {\bibinfo  {journal} {Phys. Rev. Lett.}\ }\textbf {\bibinfo {volume} {124}},\ \bibinfo {pages} {017202} (\bibinfo {year} {2020})}\BibitemShut {NoStop}%
\bibitem [{\citenamefont {Gaudet}\ \emph {et~al.}(2021)\citenamefont {Gaudet}, \citenamefont {Yang}, \citenamefont {Baidya}, \citenamefont {Lu}, \citenamefont {Xu}, \citenamefont {Zhao}, \citenamefont {Rodriguez-Rivera}, \citenamefont {Hoffmann}, \citenamefont {Graf}, \citenamefont {Torchinsky}, \citenamefont {Nikoli{\'{c}}}, \citenamefont {Vanderbilt}, \citenamefont {Tafti},\ and\ \citenamefont {Broholm}}]{Gaudet2021}%
  \BibitemOpen
  \bibfield  {author} {\bibinfo {author} {\bibfnamefont {J.}~\bibnamefont {Gaudet}}, \bibinfo {author} {\bibfnamefont {H.-Y.}\ \bibnamefont {Yang}}, \bibinfo {author} {\bibfnamefont {S.}~\bibnamefont {Baidya}}, \bibinfo {author} {\bibfnamefont {B.}~\bibnamefont {Lu}}, \bibinfo {author} {\bibfnamefont {G.}~\bibnamefont {Xu}}, \bibinfo {author} {\bibfnamefont {Y.}~\bibnamefont {Zhao}}, \bibinfo {author} {\bibfnamefont {J.~A.}\ \bibnamefont {Rodriguez-Rivera}}, \bibinfo {author} {\bibfnamefont {C.~M.}\ \bibnamefont {Hoffmann}}, \bibinfo {author} {\bibfnamefont {D.~E.}\ \bibnamefont {Graf}}, \bibinfo {author} {\bibfnamefont {D.~H.}\ \bibnamefont {Torchinsky}}, \bibinfo {author} {\bibfnamefont {P.}~\bibnamefont {Nikoli{\'{c}}}}, \bibinfo {author} {\bibfnamefont {D.}~\bibnamefont {Vanderbilt}}, \bibinfo {author} {\bibfnamefont {F.}~\bibnamefont {Tafti}}, \ and\ \bibinfo {author} {\bibfnamefont {C.~L.}\ \bibnamefont {Broholm}},\ }\href {\doibase 10.1038/s41563-021-01062-8} {\bibfield  {journal} {\bibinfo  {journal}
  {Nat. Mater.}\ }\textbf {\bibinfo {volume} {20}},\ \bibinfo {pages} {1650} (\bibinfo {year} {2021})}\BibitemShut {NoStop}%
\bibitem [{\citenamefont {Yao}\ \emph {et~al.}(2023)\citenamefont {Yao}, \citenamefont {Gaudet}, \citenamefont {Verma}, \citenamefont {Graf}, \citenamefont {Yang}, \citenamefont {Bahrami}, \citenamefont {Zhang}, \citenamefont {Aczel}, \citenamefont {Subedi}, \citenamefont {Torchinsky} \emph {et~al.}}]{yao2023large}%
  \BibitemOpen
  \bibfield  {author} {\bibinfo {author} {\bibfnamefont {X.}~\bibnamefont {Yao}}, \bibinfo {author} {\bibfnamefont {J.}~\bibnamefont {Gaudet}}, \bibinfo {author} {\bibfnamefont {R.}~\bibnamefont {Verma}}, \bibinfo {author} {\bibfnamefont {D.~E.}\ \bibnamefont {Graf}}, \bibinfo {author} {\bibfnamefont {H.-Y.}\ \bibnamefont {Yang}}, \bibinfo {author} {\bibfnamefont {F.}~\bibnamefont {Bahrami}}, \bibinfo {author} {\bibfnamefont {R.}~\bibnamefont {Zhang}}, \bibinfo {author} {\bibfnamefont {A.~A.}\ \bibnamefont {Aczel}}, \bibinfo {author} {\bibfnamefont {S.}~\bibnamefont {Subedi}}, \bibinfo {author} {\bibfnamefont {D.~H.}\ \bibnamefont {Torchinsky}},  \emph {et~al.},\ }\href {\doibase https://doi.org/10.1103/PhysRevX.13.011035} {\bibfield  {journal} {\bibinfo  {journal} {Phys. Rev. X}\ }\textbf {\bibinfo {volume} {13}},\ \bibinfo {pages} {011035} (\bibinfo {year} {2023})}\BibitemShut {NoStop}%
\bibitem [{\citenamefont {Drucker}\ \emph {et~al.}(2023)\citenamefont {Drucker}, \citenamefont {Nguyen}, \citenamefont {Han}, \citenamefont {Siriviboon}, \citenamefont {Luo}, \citenamefont {Andrejevic}, \citenamefont {Zhu}, \citenamefont {Bednik}, \citenamefont {Nguyen}, \citenamefont {Chen} \emph {et~al.}}]{drucker2023topology}%
  \BibitemOpen
  \bibfield  {author} {\bibinfo {author} {\bibfnamefont {N.~C.}\ \bibnamefont {Drucker}}, \bibinfo {author} {\bibfnamefont {T.}~\bibnamefont {Nguyen}}, \bibinfo {author} {\bibfnamefont {F.}~\bibnamefont {Han}}, \bibinfo {author} {\bibfnamefont {P.}~\bibnamefont {Siriviboon}}, \bibinfo {author} {\bibfnamefont {X.}~\bibnamefont {Luo}}, \bibinfo {author} {\bibfnamefont {N.}~\bibnamefont {Andrejevic}}, \bibinfo {author} {\bibfnamefont {Z.}~\bibnamefont {Zhu}}, \bibinfo {author} {\bibfnamefont {G.}~\bibnamefont {Bednik}}, \bibinfo {author} {\bibfnamefont {Q.~T.}\ \bibnamefont {Nguyen}}, \bibinfo {author} {\bibfnamefont {Z.}~\bibnamefont {Chen}},  \emph {et~al.},\ }\href {\doibase https://doi.org/10.1038/s41467-023-40765-1} {\bibfield  {journal} {\bibinfo  {journal} {Nat. Commun.}\ }\textbf {\bibinfo {volume} {14}},\ \bibinfo {pages} {5182} (\bibinfo {year} {2023})}\BibitemShut {NoStop}%
\bibitem [{\citenamefont {Suzuki}\ \emph {et~al.}(2019)\citenamefont {Suzuki}, \citenamefont {Savary}, \citenamefont {Liu}, \citenamefont {Lynn}, \citenamefont {Balents},\ and\ \citenamefont {Checkelsky}}]{suzuki2019singular}%
  \BibitemOpen
  \bibfield  {author} {\bibinfo {author} {\bibfnamefont {T.}~\bibnamefont {Suzuki}}, \bibinfo {author} {\bibfnamefont {L.}~\bibnamefont {Savary}}, \bibinfo {author} {\bibfnamefont {J.-P.}\ \bibnamefont {Liu}}, \bibinfo {author} {\bibfnamefont {J.~W.}\ \bibnamefont {Lynn}}, \bibinfo {author} {\bibfnamefont {L.}~\bibnamefont {Balents}}, \ and\ \bibinfo {author} {\bibfnamefont {J.~G.}\ \bibnamefont {Checkelsky}},\ }\href {\doibase DOI: 10.1126/science.aat0348} {\bibfield  {journal} {\bibinfo  {journal} {Science}\ }\textbf {\bibinfo {volume} {365}},\ \bibinfo {pages} {377} (\bibinfo {year} {2019})}\BibitemShut {NoStop}%
\bibitem [{\citenamefont {Kurumaji}\ \emph {et~al.}(2017)\citenamefont {Kurumaji}, \citenamefont {Nakajima}, \citenamefont {Ukleev}, \citenamefont {Feoktystov}, \citenamefont {Arima}, \citenamefont {Kakurai},\ and\ \citenamefont {Tokura}}]{kurumaji2017neel}%
  \BibitemOpen
  \bibfield  {author} {\bibinfo {author} {\bibfnamefont {T.}~\bibnamefont {Kurumaji}}, \bibinfo {author} {\bibfnamefont {T.}~\bibnamefont {Nakajima}}, \bibinfo {author} {\bibfnamefont {V.}~\bibnamefont {Ukleev}}, \bibinfo {author} {\bibfnamefont {A.}~\bibnamefont {Feoktystov}}, \bibinfo {author} {\bibfnamefont {T.-h.}\ \bibnamefont {Arima}}, \bibinfo {author} {\bibfnamefont {K.}~\bibnamefont {Kakurai}}, \ and\ \bibinfo {author} {\bibfnamefont {Y.}~\bibnamefont {Tokura}},\ }\href {\doibase https://doi.org/10.1103/PhysRevLett.119.237201} {\bibfield  {journal} {\bibinfo  {journal} {Phys. Rev. lett.}\ }\textbf {\bibinfo {volume} {119}},\ \bibinfo {pages} {237201} (\bibinfo {year} {2017})}\BibitemShut {NoStop}%
\bibitem [{\citenamefont {White}\ \emph {et~al.}(2018)\citenamefont {White}, \citenamefont {Butykai}, \citenamefont {Cubitt}, \citenamefont {Honecker}, \citenamefont {Dewhurst}, \citenamefont {Kiss}, \citenamefont {Tsurkan},\ and\ \citenamefont {Bord{\'a}cs}}]{white2018direct}%
  \BibitemOpen
  \bibfield  {author} {\bibinfo {author} {\bibfnamefont {J.~S.}\ \bibnamefont {White}}, \bibinfo {author} {\bibfnamefont {{\'A}.}~\bibnamefont {Butykai}}, \bibinfo {author} {\bibfnamefont {R.}~\bibnamefont {Cubitt}}, \bibinfo {author} {\bibfnamefont {D.}~\bibnamefont {Honecker}}, \bibinfo {author} {\bibfnamefont {C.~D.}\ \bibnamefont {Dewhurst}}, \bibinfo {author} {\bibfnamefont {L.~F.}\ \bibnamefont {Kiss}}, \bibinfo {author} {\bibfnamefont {V.}~\bibnamefont {Tsurkan}}, \ and\ \bibinfo {author} {\bibfnamefont {S.}~\bibnamefont {Bord{\'a}cs}},\ }\href {\doibase https://doi.org/10.1103/PhysRevB.97.020401} {\bibfield  {journal} {\bibinfo  {journal} {Phys. Rev. B}\ }\textbf {\bibinfo {volume} {97}},\ \bibinfo {pages} {020401} (\bibinfo {year} {2018})}\BibitemShut {NoStop}%
\bibitem [{\citenamefont {Sergienko}\ and\ \citenamefont {Dagotto}(2006)}]{sergienko2006role}%
  \BibitemOpen
  \bibfield  {author} {\bibinfo {author} {\bibfnamefont {I.~A.}\ \bibnamefont {Sergienko}}\ and\ \bibinfo {author} {\bibfnamefont {E.}~\bibnamefont {Dagotto}},\ }\href {\doibase https://doi.org/10.1103/PhysRevB.73.094434} {\bibfield  {journal} {\bibinfo  {journal} {Phys. Rev. B}\ }\textbf {\bibinfo {volume} {73}},\ \bibinfo {pages} {094434} (\bibinfo {year} {2006})}\BibitemShut {NoStop}%
\bibitem [{\citenamefont {Nikoli{\'{c}}}(2020)}]{Nikolic2020a}%
  \BibitemOpen
  \bibfield  {author} {\bibinfo {author} {\bibfnamefont {P.}~\bibnamefont {Nikoli{\'{c}}}},\ }\href {\doibase 10.1103/PhysRevB.102.075131} {\bibfield  {journal} {\bibinfo  {journal} {Phys. Rev. B}\ }\textbf {\bibinfo {volume} {102}},\ \bibinfo {pages} {075131} (\bibinfo {year} {2020})}\BibitemShut {NoStop}%
\bibitem [{\citenamefont {Nikoli{\'{c}}}(2021)}]{Nikolic2020b}%
  \BibitemOpen
  \bibfield  {author} {\bibinfo {author} {\bibfnamefont {P.}~\bibnamefont {Nikoli{\'{c}}}},\ }\href {\doibase 10.1103/PhysRevB.103.155151} {\bibfield  {journal} {\bibinfo  {journal} {Phys. Rev. B}\ }\textbf {\bibinfo {volume} {103}},\ \bibinfo {pages} {155151} (\bibinfo {year} {2021})}\BibitemShut {NoStop}%
\bibitem [{\citenamefont {Ozawa}\ \emph {et~al.}(2016)\citenamefont {Ozawa}, \citenamefont {Hayami}, \citenamefont {Barros}, \citenamefont {Chern}, \citenamefont {Motome},\ and\ \citenamefont {Batista}}]{Ozawa2016}%
  \BibitemOpen
  \bibfield  {author} {\bibinfo {author} {\bibfnamefont {R.}~\bibnamefont {Ozawa}}, \bibinfo {author} {\bibfnamefont {S.}~\bibnamefont {Hayami}}, \bibinfo {author} {\bibfnamefont {K.}~\bibnamefont {Barros}}, \bibinfo {author} {\bibfnamefont {G.-W.}\ \bibnamefont {Chern}}, \bibinfo {author} {\bibfnamefont {Y.}~\bibnamefont {Motome}}, \ and\ \bibinfo {author} {\bibfnamefont {C.~D.}\ \bibnamefont {Batista}},\ }\href {\doibase 10.7566/JPSJ.85.103703} {\bibfield  {journal} {\bibinfo  {journal} {J. Phys. Soc. Jpn}\ }\textbf {\bibinfo {volume} {85}},\ \bibinfo {pages} {103703} (\bibinfo {year} {2016})}\BibitemShut {NoStop}%
\end{thebibliography}%

\noindent

\end{document}